\documentclass{article}
\usepackage{sty/ijcai26}

\usepackage{times}
\usepackage{soul}
\usepackage{url}
\usepackage[hidelinks]{hyperref}
\usepackage[utf8]{inputenc}
\usepackage[small]{caption}
\usepackage{graphicx}
\usepackage{algorithm}
\usepackage[noEnd=true,indLines=true,commentColor=black]{sty/algpseudocodex}
\usepackage{amsmath,amssymb,amsthm}
\usepackage{siunitx}
\usepackage{booktabs}
\usepackage{hyperref}

\makeatletter
\AddToHook{env/algorithmic/before}{%
  \def\@currentcounter{ALG@line}%
}
\makeatother
\usepackage{cleveref}
\crefalias{ALG@line}{line}

\usepackage[switch]{lineno}
\usepackage{sty/custom}

\title{From Gridworlds to Warehouses:\\
  Adapting Lightweight One-shot Multi-Agent Pathfinding for AGVs
}
\author{
  Hiroki Nagai$^{1,2}$
  \and
  Keisuke Okumura$^1$\\
  \affiliations
  $^1$National Institute of Advanced Industrial Science and Technology (AIST), Japan\\
  $^2$Keio University, Japan
  \emails
  \{nagai.hiroki39,okumura.k\}@aist.go.jp
}

\begin{document}
\maketitle
\begin{abstract}
Multi-agent pathfinding (MAPF) under one-shot planning is a core component of warehouse automation, yet classical formulations typically assume four-connected 2D grids with unit-time moves in four directions. 
To fill reality gaps while still being trackable with discrete combinatorial search, this work proposes a more practical counterpart tailored to differential-drive AGVs.
We term this \emph{multi-agent warehouse pathfinding (MAWPF)}, featured with four constraints:
\emph{(i)}~agent actions are restricted to straight motion and in-place rotation; \emph{(ii)}~rotations require multi-step costs;
\emph{(iii)}~acceleration and deceleration are considered, and;
\emph{(iv)}~follower collisions are prohibited to prevent rear-end crashes.
To solve MAWPF efficiently, we adapt representative suboptimal MAPF algorithms---PP, LNS2, PIBT, and LaCAM---and conduct comprehensive benchmarking. 
Our experiments reveal that PP and LNS2 struggle to solve instances with many agents, while PIBT-based approaches achieve preferable scalability with increased solution cost.
We believe that these constitute an important step toward adapting classical gridworld MAPF to operational warehouse setups.
\end{abstract}

\section{Introduction}
Multi-agent pathfinding (MAPF) is a versatile abstraction for a variety of multi-agent planning problems, such as traffic intersection management~\cite{traffic}, railway scheduling~\cite{railway}, autonomous parking~\cite{parking}, and conveyor routing~\cite{kato2026conveyor}. To enhance its engineering transferability beyond specific applications or infrastructures, most academic studies on MAPF adopt a gridworld scenario~\cite{gridworld}, in which all agents use a planar graph as a workspace representation—typically a four-connected 2D grid—and take unit-length actions in unit time, in a fully synchronous manner. In this gridworld context, at each timestep, each agent occupies exactly one cell, which simplifies collision management. Such an assumption indeed reflects several characteristics of real-world \emph{warehouse} setups~\cite{warehouse}, a representative MAPF application, where autonomous guided vehicles (AGVs) perform tasks while following predefined grid structures.

{
  \begin{figure}[t!]
    \centering
    \includegraphics[width=1.0\linewidth]{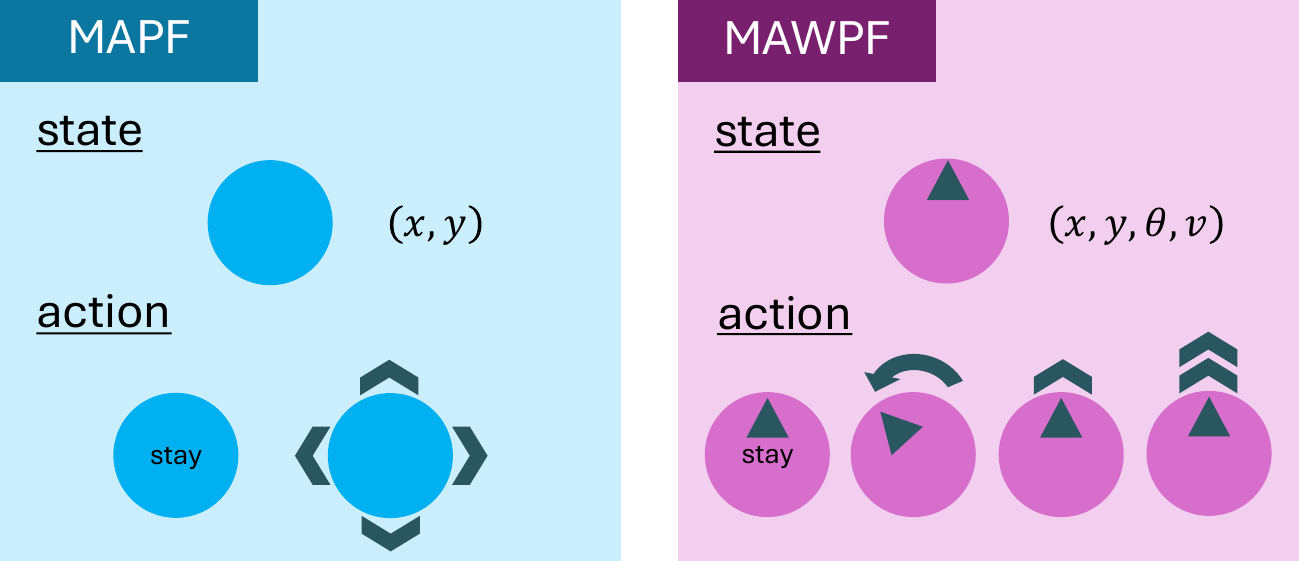}
    \caption{Classical MAPF and MAWPF.}
    \label{fig:MAWPF}
  \end{figure}
}

This gridworld abstraction has enabled the research community to develop a wide variety of capable and foundational MAPF algorithms. In fact, recent developments of real-time, scalable MAPF algorithms---such as PIBT~\cite {PIBT} and LaCAM~\cite{LaCAM}---are remarkable, capable of handling hundreds of agents in a second. Meanwhile, to deploy these algorithms in actual warehouse setups and maximize overall system performance, it is necessary to incorporate robot kinodynamic constraints tailored to AGVs. Such constraints include \textbf{\emph{(i)} rotation constraints}, where an AGV cannot move in an arbitrary direction, \textbf{\emph{(ii)} turning action costs}, where an AGV cannot complete a 90-degree rotation within a single time unit, \textbf{\emph{(iii)} follower conflicts}, where an AGV must avoid entering a cell that was just vacated, and  \textbf{\emph{(iv)} acceleration and deceleration considerations}, where an AGV can traverse multiple cells within a single time unit.

Although several established post-processing methods exist to translate gridworld MAPF plans into AGV-compatible ones~\cite{AGV}, these approaches inherently rely on suboptimal plans due to their use of over-simplified robot dynamics, making it difficult to achieve high system performance~\cite{yan2025analyzing}.
Several studies have also explored extensions of gridworld MAPF---for example, incorporating rotation constraints~\cite{rotation,chan2024league}---yet a unified formulation that integrates multiple practical AGV-specific considerations within a single framework is missing.

The aforementioned four AGV-related considerations are common in multi-AGV systems across different infrastructures. 
Given their economic impact and universality, we consider it valuable to define the corresponding problem that bridges academic research and industrial practice in MAPF.

To this end, we introduce the \emph{multi-agent warehouse pathfinding (MAWPF)} problem as a gridworld counterpart of MAPF tailored to warehouse environments, and investigate how computationally lightweight MAPF algorithms can be adapted to this setting.
MAWPF is carefully designed to capture the aforementioned four constraints stemming from AGV dynamics, while discrete, combinatorial search-based MAPF algorithms remain applicable with modest adaptation.
In particular, we describe how representative MAPF solvers---Prioritized Planning (PP)~\cite{PP}, LNS2~\cite{LNS2}, PIBT, and LaCAM---are applied to MAWPF.
We then systematically evaluate these solvers to delineate their current capabilities and limitations.
We believe that this effort represents an important step toward upgrading conventional MAPF benchmarking with real-world grounding and operational relevance.

We begin the rest with the classical MAPF formulation and then define MAWPF, followed by reviews of MAPF algorithms and kinematics-aware variants.
After that, we describe how MAPF algorithms are applied to MAWPF and report their empirical performance.
The code and appendix are available from \url{https://github.com/hirokiNagai-39/mawpf}.

\section{Classical MAPF}
To facilitate understanding of MAWPF, we here describe the classical yet most commonly-used MAPF formulation~\cite{gridworld} to date.
The system consists of a set of agents $A = \{1, 2, \ldots, n\}$ operating on a graph $G=(V, E)$ under a discrete, globally synchronized time model.
At each timestep, an agent may either wait at its current vertex or move to an adjacent vertex. 
Feasibility is defined with respect to two collision constraints: \emph{(i)~vertex conflicts}, where two agents occupy the same vertex at the same time, and \emph{(ii)~edge conflicts}, where two agents traverse the same edge in opposite directions within a single timestep. 
Given distinct start and goal vertices $(s_i,g_i)\in V\times V$ for each agent $i\in A$, the one-shot MAPF problem asks for a finite sequence of actions for every agent that brings all agents from their starts to their respective goals without conflicts. 
Solution quality is evaluated using the \emph{sum-of-costs (SoC)} (aka. flowtime), defined as the sum, over all agents, of the time until each agent first reaches its goal and remains there thereafter.

\section{MAWPF}
Classical MAPF provides a simple and intuitive abstraction for coordinating multiple agents, which has made it a widely used benchmark problem.
However, it often departs from the operational realities of warehouse transportation, where differential-drive AGVs are prevalent and their motion is constrained by heading changes and speed control.
To investigate how standard MAPF algorithms behave under such more realistic conditions, we first introduce a problem formulation that reflects these constraints.
Specifically, we propose \emph{multi-agent warehouse pathfinding (MAWPF)} as an MAPF tailored to the realities of automated warehouse distribution.
The differences from classical MAPF are described below, and \cref{fig:MAWPF} provides an overview.

\subsection{Problem Definition}

\paragraph{Workspace.}
We consider a four-connected 2D grid map with \emph{static} obstacles.
The workspace is modeled as a graph $G=(V, E)$, where $V$ is the set of obstacle-free cells and $E$ connects pairs of four-neighbor cells; agents move along edges in $E$.

\paragraph{Configuration.}
In MAWPF, a \emph{configuration} $\Q$ refers to the array of states of all agents at a given timestep.
Unlike classical MAPF, each agent state includes not only its grid location but also its \emph{direction} and \emph{speed}. Formally,
\begin{equation}
\Q[i]=(x_i,y_i,\theta_i,v_i)\quad v_i\in \{0,1,\ldots,V_{max}\}.    
\label{eq:MAWPFconfig}
\end{equation}
Here, speed $v$ is the number of grid cells advanced in the next time step, and $\theta$ is the angle measured counterclockwise from the positive $x$-axis.
In addition, the model is parameterized by $V_{\max}$, the maximum number of grid cells an agent can traverse in one timestep, and $T_{\mathrm{rot}}$, the number of timesteps required to complete a \SI{90}{\deg} rotation.

\paragraph{Action.}
At each timestep, agent $i$ updates its state by applying \emph{(i)~movement} and then \emph{(ii)~speed change}.
In the movement phase, the agent selects exactly one of the following: \emph{stay}, \emph{forward}, or \emph{rotation}.
The \emph{stay} action keeps the agent at the current cell and is available only when $v_i=0$.
The \emph{forward} action moves the agent straight ahead by $v_i$ cells in the direction $\theta_i$; this action is available only when
$\theta_i \in \{0,90,180,270\}\,\deg$.
The \emph{rotation} action rotates the agent on the spot by $90/T_{\mathrm{rot}}\,\deg$ clockwise or counterclockwise, and it is available only when $v_i=0$.

After the movement phase, the agent performs a speed change by selecting exactly one of \emph{keep}, \emph{acceleration}, or \emph{deceleration}.
The \emph{keep} action leaves the speed unchanged.
The \emph{acceleration} action increases the speed by one, i.e., $v_i \leftarrow v_i+1$, and is available only when $v_i < V_{\max}$ and
$\theta_i \in \{0,90,180,270\}\,\deg$.
The \emph{deceleration} action decreases the speed by one, i.e., $v_i \leftarrow v_i-1$, and is available only when $v_i>0$ and
$\theta_i \in \{0,90,180,270\}\,\deg$.
Consequently, an agent cannot accelerate or stop abruptly. 

\paragraph{Vertex Occupation.}
At timestep $t$, during the movement $\Q_{t}[i]{\rightarrow}\Q_{t+1}[i]$, agent $i$ occupies the entire line segment range from geometric vertex $(x_i^t,y_i^t)$ to $(x_i^{t+1},y_i^{t+1})$:
e.g., when an agent with $v_i=2$ moves from $(0,0)$ to $(2,0)$, it occupies the three geometric vertices $(0,0)$, $(1,0)$, and $(2,0)$.

\paragraph{Collision.}
At a given time step, a collision is considered to occur when a geometric vertex $(x,y)$ is occupied by a pair of agents $i, j \in A$, $i \neq j$.
This collision definition encompasses not only \emph{vertex collisions} and \emph{edge collisions}, but also \emph{follower collisions}.
A follower collision is a constraint preventing agents from entering the grid cells occupied by other agents in the previous timestep, in case those agents malfunctioned and did not move during that step.

\paragraph{MAWPF problem.}
Let $\S$ and $\G$ denote the start and goal configurations, respectively. The solution to MAWPF is a sequence of collision-free configurations $\Pi=(\Q_0,\ldots,\Q_k)$ that satisfy $\Q_0=\S$ and $\Q_k=\G$.

\paragraph{Problem Difficulty.}
While it is possible to comprehend MAPF and MAWPF as pathfinding on a graph whose vertices are the configurations, a key difference between the MAWPF graph and the one used in classical MAPF is that the MAWPF graph is \emph{directed}.
This directionality arises from the assumption that agents cannot accelerate or decelerate abruptly: once an agent commits to a motion state, the set of feasible next states is constrained, and an agent is not guaranteed to be able to return to its previous vertex.
As a result, the planner can easily enter a stuck (dead-end) situation where no feasible continuation exists without violating kinematic feasibility or collision constraints.
With multiple agents, such irreversibility increases the likelihood of system-wide deadlocks, making planning failures more frequent.
In these senses, solving MAWPF is inherently more difficult than solving classical MAPF on undirected graphs where agents can more readily backtrack and resolve local conflicts.

\section{Related Work}
Having defined MAWPF as a warehouse-grounded extension of classical MAPF, we next review prior work that informs our study from two perspectives: scalable MAPF solvers developed under the classical formulation, and MAPF variants that explicitly incorporate kinodynamic constraints.
\paragraph{MAPF Solvers.}
A naive approach to solve MAPF optimally is to employ joint-state search over configurations via A$^\ast$~\cite{Astar,standley2010finding}.
However, it is impractical because the branching factor grows exponentially with the number of agents, quickly making the search intractable.
This motivates researchers to explore a different style of planning representation; for example, CBS~\cite{CBS}, a celebrated MAPF algorithm, searches over constraint sets and replans agent-wise paths to resolve collisions.

CBS and its extension (e.g., \cite{li2021eecbs}) is more scalable than joint A$^\ast$, yet it can still miss tight time budgets in scenarios with hundreds of agents or more.
Therefore, a substantial line of research has focused on \emph{unbounded suboptimal but scalable} MAPF algorithms that sacrifice some degree of optimality to achieve real-time performance on large instances.
Representative lightweight methods include \emph{Prioritized Planning (PP)}~\cite{PP}, \emph{LNS2}~\cite{LNS2}, \emph{PIBT}~\cite{PIBT}, and \emph{LaCAM}~\cite{LaCAM}; these will be reviewed in \cref{sec:algo}.
In particular, the combination of LaCAM with PIBT has been reported as highly effective, enabling genuinely large-scale planning (on the order of thousands to even ten-thousands of agents) while maintaining practical runtime.

In this paper, we build on these scalable, lightweight MAPF techniques rather than relying on optimal algorithms since our target application is large-scale warehouse transportation.
Moreover, the increased problem complexity arising from advanced kinodynamic constraints makes optimal planning less tractable, thereby reinforcing the practical relevance of suboptimal approaches.

\paragraph{Kinodynamic-considered MAPF.}
Classical MAPF typically abstracts away robot kinodynamics by assuming holonomic, discrete-time motion with instantaneous turns and instantaneous changes of velocity; however, for real AGVs in warehouses, such simplifications can misrepresent the true feasibility and execution cost.

Two approaches are commonly adopted to address this mismatch.
The first category post-processes gridworld MAPF solutions to make them compatible with AGV dynamics.
For example, \cite{AGV} employs a simple temporal network to translate kinematics-agnostic plans into ones that respect non-holonomic motion with bounded translational and rotational velocities, while \cite{yan2025multi} uses a continuous-time single-agent search to obtain such plans.
This direction benefits from the scalability of grid-based solvers; however, the resulting solutions are often suboptimal, as high-level grid-based plans inherently introduce discrepancies between estimated and actual execution costs~\cite{yan2025analyzing}.

Another line of work extends the classical MAPF formulation to account for robot dynamics explicitly and solves the resulting problem directly.
This approach narrows the reality gap and enables the computation of plans with improved execution fidelity.
For example, MAPF with Turn Actions (MAPF$_T$)~\cite{rotation} augments the action space by modeling turning as an explicit action, while MAPF with Kinematic Constraints (MAPFKC)~\cite{SIPP} incorporates speed and acceleration, and CL-MAPF~\cite{CLMAPF} focuses on Ackermann-steering, car-like robots, to name just a few.
Our MAWPF falls within this category, but two features distinguish it in terms of the practicality and applicability of existing MAPF methods:
\emph{(i)}~MAWPF adopts a fully discrete state space, which allows existing discrete search-based MAPF algorithms to be adapted directly (unlike MAPFKC);
\emph{(ii)}~MAWPF jointly considers AGV-specific constraints, including motion, rotation, and collision characteristics (unlike MAPF$_T$ and CL-MAPF).

\section{Algorithms}
\label{sec:algo}

This section describes how we adapt existing lightweight MAPF algorithms to solve MAWPF under our differential-drive action model and warehouse-specific safety constraints. 

\subsection{Prioritized Planning (PP)}
PP~\cite{PP} is a sub-optimal and incomplete solver that assigns priorities to agents and then plans collision-free paths in priority order.
Similar to MAPF implementation,
PP for MAWPF plans agents sequentially with a fixed priority ordering. For each agent, we run space--time A*~\cite{silver2005cooperative} on the \emph{MAWPF configuration graph}, whose vertices encode kinodynamic configurations and whose edges correspond to admissible MAWPF actions (e.g., forward motion and in-place rotation). Previously planned trajectories are treated as dynamic obstacles via a time-indexed reservation table: a transition at time $t$ is rejected if it occupies any reserved cell at layer $t$. Since one MAWPF action may traverse multiple cells, we reserve and check all cells along the swept segment. The resulting path is then committed to the reservation table, including goal-hold after arrival.
Overall, this algorithm constitutes the simplest baseline approach for solving MAWPF.

\subsection{MAPF Large Neighborhood Search (LNS2)}
LNS2~\cite{LNS2} is a sub-optimal and incomplete solver that first computes each agent’s path without considering collisions, and then resolves collisions via large neighborhood search.
LNS2 for MAWPF is a destroy-and-repair local search that reduces collisions by replanning only a subset of agents. It starts with \emph{soft} prioritized planning: \emph{as in PP for MAWPF}, each agent is planned on the MAWPF state graph with a time-indexed reservation structure, but collisions are allowed with an extra penalty proportional to the number of reserved cells intersected by the swept motion segment. It then repeatedly selects a subset of agents and destroys and repairs their paths until collisions are eliminated or the time budget is exhausted.

\subsection{PIBT}
PIBT~\cite{PIBT} maps a collision-free configuration $\Q\from \in V^{|A|}$ to another $\Q\to$; iterating this mapping yields an MAPF solver that is incomplete and sub-optimal. Agents are assigned \emph{priorities} and reserve vertices for the next timestep in priority order according to their \emph{preferences}. If a low-priority agent conflicts with a high-priority agent, it inherits the higher-priority agent’s position, recursively calls PIBT to vacate the cell, and then reserves another vertex.

\paragraph{Multi-step PIBT.}
Under a differential-drive model with only \emph{move forward} and \emph{rotate in place}, an agent receiving priority inheritance cannot always vacate immediately: because it cannot move laterally, side interactions may require multiple primitive actions before the contested cell is cleared. As shown in \cref{tab:pibt_p1_p6_success_rate}, the original PIBT can hardly solve MAWPF. We therefore extend PIBT to a \emph{multi-step} variant, allowing an inherited agent to execute a short action sequence (e.g., rotate then move forward) to vacate in a kinematically feasible way, enabling PIBT-style coordination for differential-drive MAPF. Along this line, Enhancing PIBT~\cite{EPIBT} solves lifelong MAPF for differential two-wheeled AGVs by reserving short-horizon segments rather than a single vertex; we similarly develop a multi-step PIBT-based algorithm for one-shot MAWPF, detailed at the end of this section.

\paragraph{Definition of Stop Path.}
In original PIBT, if agent $i$'s plan fails, it chooses \emph{stay}; in MAWPF, emergency stops are impossible, so \emph{stay} is not always selectable. We thus define the \emph{stop path} (the counterpart of \emph{stay}) as the shortest path from the current state to a stop. For $L=3$, examples include:
$(0,0,0,3)\rightarrow (3,0,0,2)\rightarrow (5,0,0,1)\rightarrow (6,0,0,0)$.
When agent $i$'s plan fails, it is assigned the \emph{stop path}.

\paragraph{Priority Inheritance for MAWPF.}
While several inheritance rules are possible, using the \emph{stop path} as the analogue of \emph{stop} makes the following rule natural: as shown in \cref{fig:multistepPIBT}, among agents not yet assigned a path, priority is inherited by the agent whose stop path conflicts with agent $i$.

\paragraph{MAWPF Adaptation.}
The blue parts in \cref{alg:pibtMAWPF} indicate the modifications from the original PIBT.
Starting from the current configuration $\Q\from$, the procedure constructs an $L$-length configuration sequence $\{\Q\to_0,\Q\to_1,\ldots,\Q\to_{L-1}\}$ by assigning each agent $i\in A$ a feasible horizon path; agents with $\Q\to_{L-1}[i]{=}\bot$ are planned via the recursive procedure $\PIBT(i)$ (Lines~1--3). We enumerate all admissible $L$-step action sequences via Breadth-First Search on the MAWPF configuration graph. Inside $\PIBT(i)$, candidate horizon paths $Paths\gets\pathgen(\Q\from[i])$ are sorted by the distance from $Path[L-1]$ to $g_i$ (Lines~8--9), and tested in that order: the algorithm checks whether reserving the entire segment causes any collision (Lines~10--12) and commits a feasible candidate by setting $\Q\to[i]\gets Path$ (Line~13). It then applies segment-level \emph{priority inheritance}: for each unplanned agent $j$, it checks whether its default $StopPath[j]$ would collide with current reservations (Procedure \Inherit, Lines~4--6; condition in Lines~15); if so, it calls $\PIBT(j)$, and if this fails it backtracks and rejects the current candidate (Lines~15--16). If no candidate yields consistent reservations, it assigns $\Q\to[i]\gets StopPath[i]$ and returns \textsc{INVALID} (Lines~20--21).

\paragraph{Algorithm Improvements.}
PIBT for MAWPF incorporates implementation-level refinements to improve scalability under kinodynamic constraints.
\begin{itemize}
    \item \textbf{Division sort} (\cref{algo:pibtMAWPF:sort}):
    Instead of fully sorting $Paths$, we repeatedly sort only the top-$K$ candidates and, if none is adopted, proceed to the next top-$K$ until a path is selected.
    In practice, substantially fewer than $\lvert Paths \rvert$ elements are often sorted, yielding a noticeable reduction in computation.

    \item \textbf{Pruning} (\cref{algo:pibtMAWPF:getPaths}):
    Among length-$L$ candidate paths with identical first and last configurations, we keep only the one with the smallest number of moves, removing redundant candidates that return to the same endpoint configuration and discouraging oscillatory behavior.
    By reducing $\lvert Paths \rvert$, pruning directly accelerates a single configuration-generation call.
\end{itemize}

\begin{table}[t]
  \centering
  
  \begin{tabular}{lcccc}
    \toprule
     & $|A|{=}5$ & $10$ & $15$ & $20$ \\
    \midrule
    PIBT & 0.76  & 0.20   & 0.04   & 0.00  \\
    Multi-step PIBT ($L{=}6$)& 1.00  & 1.00 & 1.00  & 1.00  \\
    \bottomrule
  \end{tabular}

  \caption{Success rate comparison between original PIBT and multi-step PIBT for MAWPF on \emph{random-64-64-20} with $V_{\mathrm{max}}{=}2, T_{\mathrm{rot}}{=}2$, using the rolling horizon method.}
  \label{tab:pibt_p1_p6_success_rate}
\end{table}

{
  \begin{figure}[th!]
    \centering
    \includegraphics[width=1\linewidth]{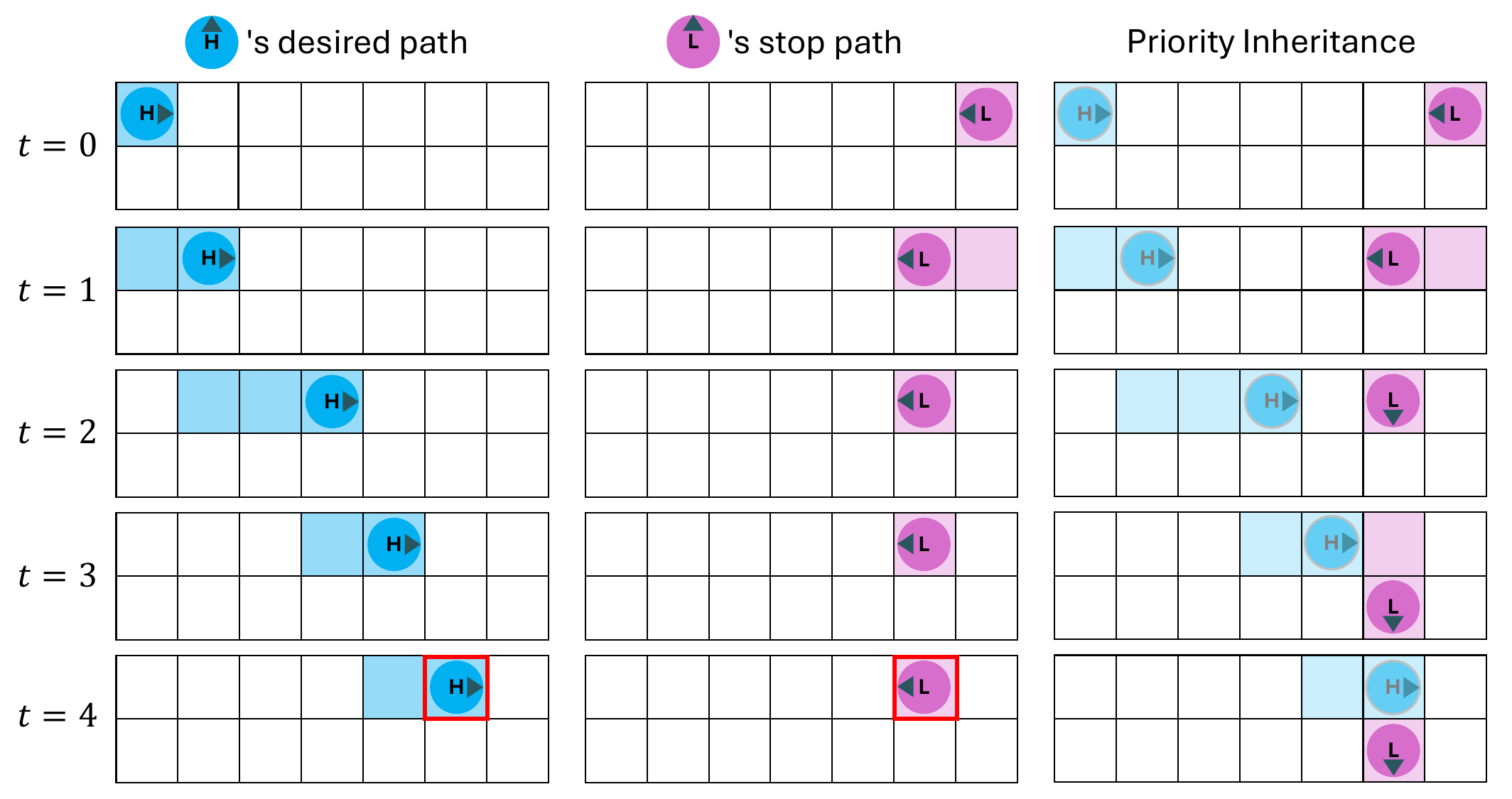}
    \caption{Priority Inheritance for MAWPF. Illustrates case $L=4$. Colored vertices represent those occupied by agents at each time step. The path of the higher-priority agent (left) and the stopping path of the lower-priority agent (center) collide at $t=4$, triggering priority inheritance. The lower-priority agent is then planned (right).}
    \label{fig:multistepPIBT}
  \end{figure}
}

%--------------------------------------
% Algorithm 4: PIBT for MAWPF
%--------------------------------------
\begin{algorithm}[t]
  \caption{PIBT for MAWPF}
  \label{alg:pibtMAWPF}
  \begin{algorithmic}[1]
  \small
    \Input{configuration $\Q\from$, agents $A$, goals $(g_i)^{i \in A}$}
    \Output{configurations $\diff{\{\,\Q\to_0,\Q\to_1,...,\Q\to_{L-1}\,\}}$}
    \For{$i \in A$}
      \If{$\Q\to_{L-1}[i] = \bot$}  $\PIBT(i)$
      \EndIf
    \EndFor
    \State \Return $\diff{\{\,\Q\to_0,\Q\to_1,...,\Q\to_{L-1}\,\}}$

    \Procedure{\Inherit}{$j$}
    \IfSingle{$StopPath[j]$ would cause a collision}{\Return \true}
    \State \Return \false
    \EndProcedure

    \Procedure{\PIBT}{$i$} 
      \State $\diff{Paths \gets \pathgen(\Q\from[i])}$ 
      \label{algo:pibtMAWPF:getPaths}      
      %\Comment {Enumerate paths of path_length steps}
      \State sort $Paths$ in ascending order of
      \label{algo:pibtMAWPF:sort}
      \Statex $\dist(\diff{Path[L-1]}, g_i)$ for $Path \in Paths$

      \For{$Path \in Paths$}
        \State $collision \gets \false$
        \If{$Path$ would cause a collision}
        \Continue
        \EndIf
        
        \State $\diff{\Q\to{[i]} \gets Path}$
        \For{$j \in A$}
        \If{$j \neq i 
        \land \Q\to_{L-1}[j] = \bot \land \diff{\Inherit(j)}$}
          \If{$\PIBT(j) = \invalid$}
            \State $collision \gets \true$
            ; \Break
          \EndIf
        \EndIf

        \EndFor

       \If{$collision$}
       \Continue
       \EndIf

        \State \Return \valid
      \EndFor

      \State $\diff{\Q\to[i] \gets Stop Path[i]}$
      \State \Return \invalid
    \EndProcedure
  \end{algorithmic}
\end{algorithm}

\begin{table}[t]
  \centering
  
  \begin{tabular}{lcccc}
    \toprule
     & $|A|{=}50$ & $100$ & $150$ & $200$ \\
    \midrule
    Macro-successor & 0.00 & 0.00 & 0.00 & 0.00 \\
    Rolling horizon & 1.00 & 1.00 & 1.00 & 1.00 \\
    \bottomrule
  \end{tabular}

  \caption{Effect of rolling horizon (\emph{random-64-64-20}, $L{=}6$, $V_{\mathrm{max}}{=}2$, $T_{\mathrm{rot}}{=}2$).}
  \label{tab:rh}
\end{table}

\subsection{LaCAM}
LaCAM~\cite{LaCAM} is a search-based meta-algorithm for MAPF that rapidly finds feasible solutions on large, dense instances.
It reduces branching by avoiding explicit enumeration of the joint action space and instead generating successors \emph{lazily} using constraints.
LaCAM is \emph{complete}: it returns a solution if one exists and otherwise reports failure.
To do so, LaCAM employs two level search.
The \emph{high level} searches over \emph{configurations}, constructing a sequence from $\S$ to $\G$.
The \emph{low level} grows a \emph{constraint tree} whose nodes impose requirements such as ``agent $i$ must be at vertex $v$ in the next configuration.''
For a selected node, LaCAM invokes a configuration generator such as PIBT to produce a valid next configuration; if generation fails, it switches to another node, by lazy successor enumeration.

\paragraph{MAWPF Adaptation.}
We retain the high-level search and replace the configuration generator with \emph{PIBT for MAWPF} (\cref{alg:pibtMAWPF}).
As this generator returns an $L$-step configuration array, two integration options arise.
One treats the $L$-step array as a \emph{macro-successor}, advancing the high level by $L$ timesteps per expansion.
The other uses only the first configuration and advances in a \emph{rolling-horizon} manner, committing one timestep and re-invoking the generator at the next node.

Our evaluation shows that the macro-successor scheme fails to solve MAWPF instances (\cref{tab:rh}); thus, we adopt the rolling-horizon integration.
This is likely due to the directed MAWPF configuration graph, where committing to multi-step expansions can quickly lead to dead-end configurations.
Further details are provided in the appendix.
Following this observation, we also adopt the rolling-horizon scheme when using \emph{PIBT for MAWPF} alone.

\section{Evaluation}
This section evaluates the performance of the MAWPF algorithms: \textbf{PP}, \textbf{LNS2}, \textbf{PIBT}, and \textbf{LaCAM}.

\subsection{Experimental Setup}
The evaluation metrics are planning \emph{success rate}, \emph{runtime}, and \emph{sum-of-cost}.
Experiments were conducted on a laptop equipped with an M3 Pro Apple Silicon chip and \SI{18}{\giga\byte} of RAM. All code was written in C++.
Unless specified, $V_{\max}$ and $T_{\mathrm{rot}}$ were set to $2$.
For PIBT, we varied the lookahead horizon $L$ from $5$ to $7$.
The evaluation used 12 maps from the MAPF benchmark~\cite{gridworld}; see \cref{fig:oneshot-mapf}.
The values reported were generated from 25 test cases, prepared for each map and each number of agents, with randomized start and goal states.
In addition to positions, each start/goal includes a heading sampled uniformly from $\{0,90,180,270\}\,\deg$, and the agent speed at both the start and the goal is set to $0$.

{
\newcommand{\entry}[1]{
  \begin{minipage}{0.1545\linewidth}
    \centering
    
    \includegraphics[width=1\linewidth,trim={0.34cm 0cm 0.28cm 0cm},clip]{fig/raw/result_fig3/#1_results.pdf}
  \end{minipage}
}

\begin{figure*}[t!]
  \centering
  \entry{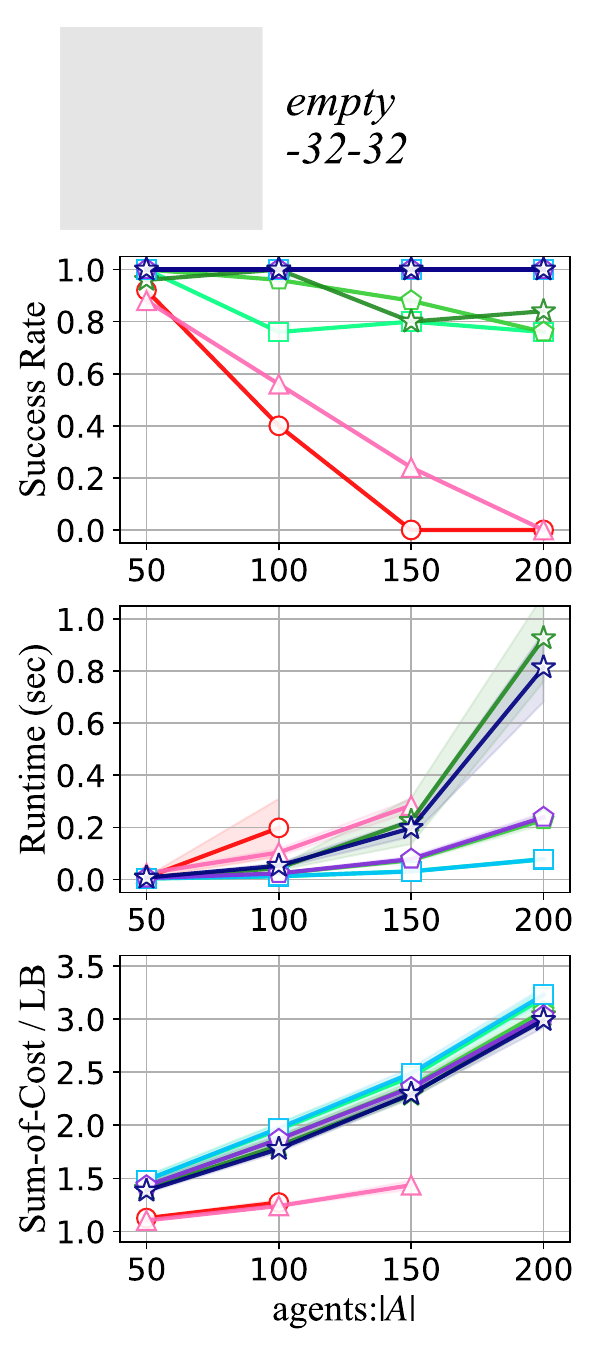}
  \entry{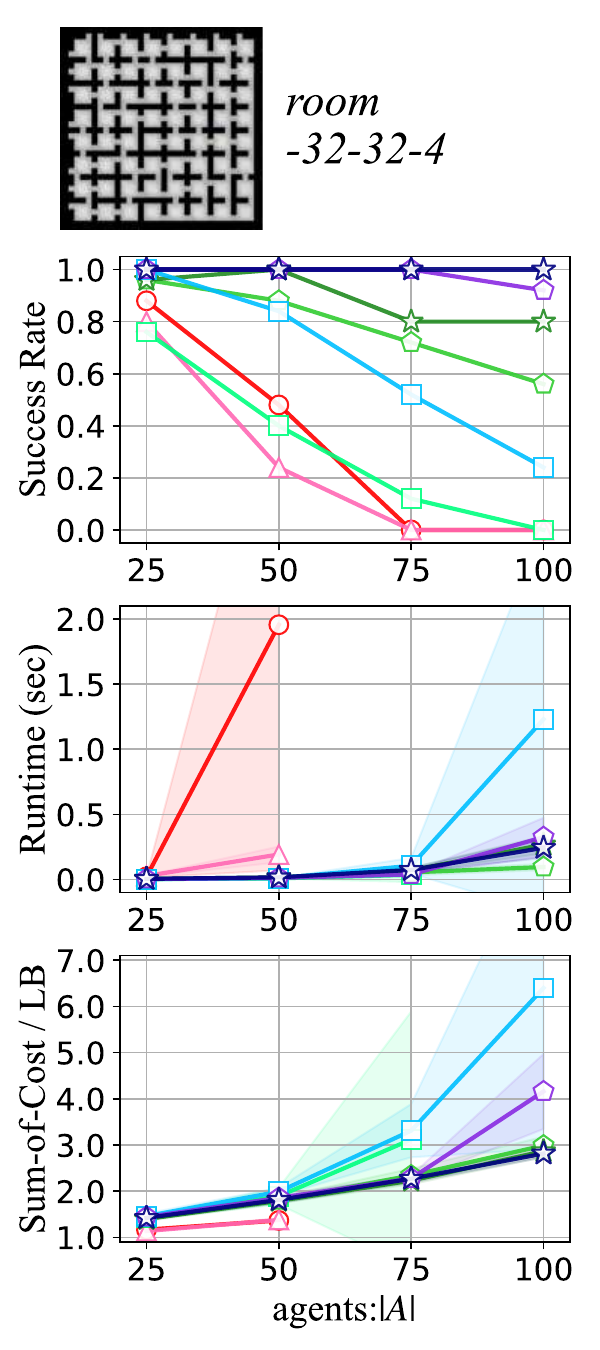}
  \entry{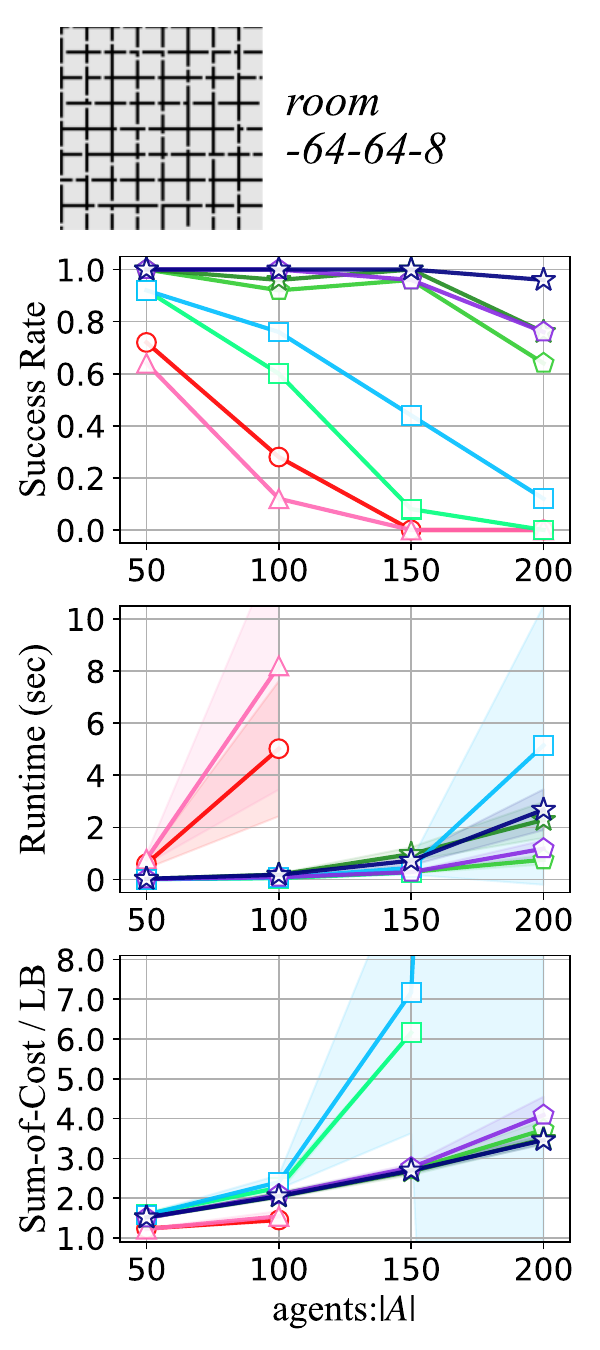}
  \entry{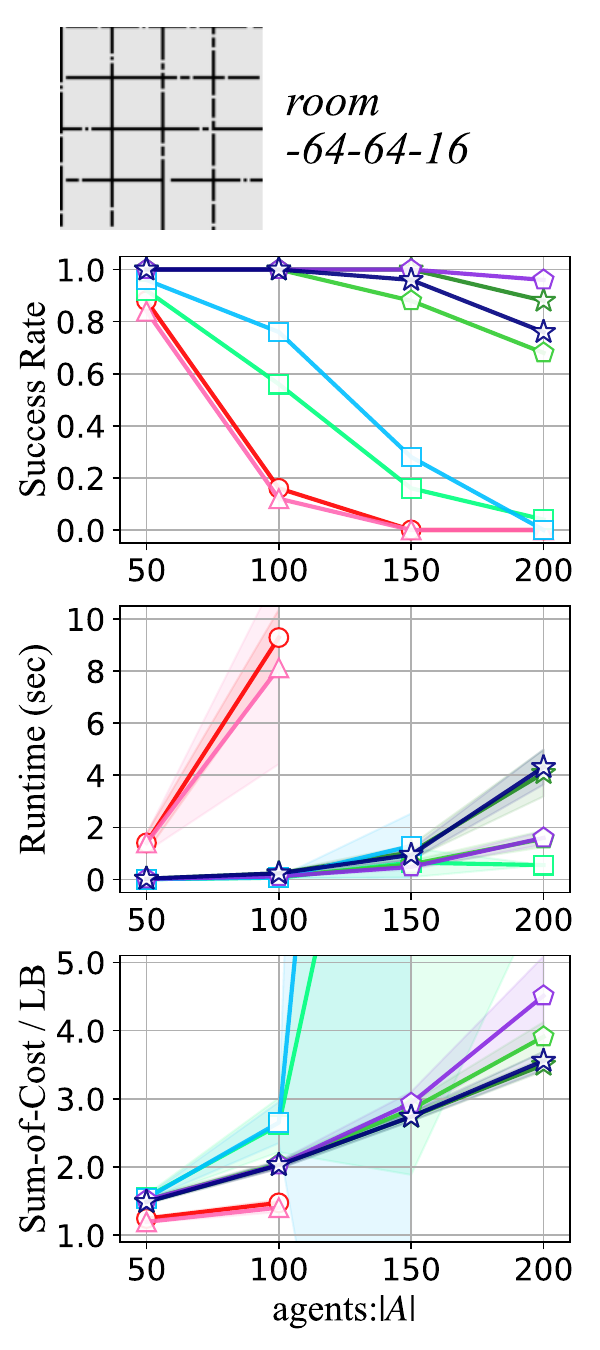}
  \entry{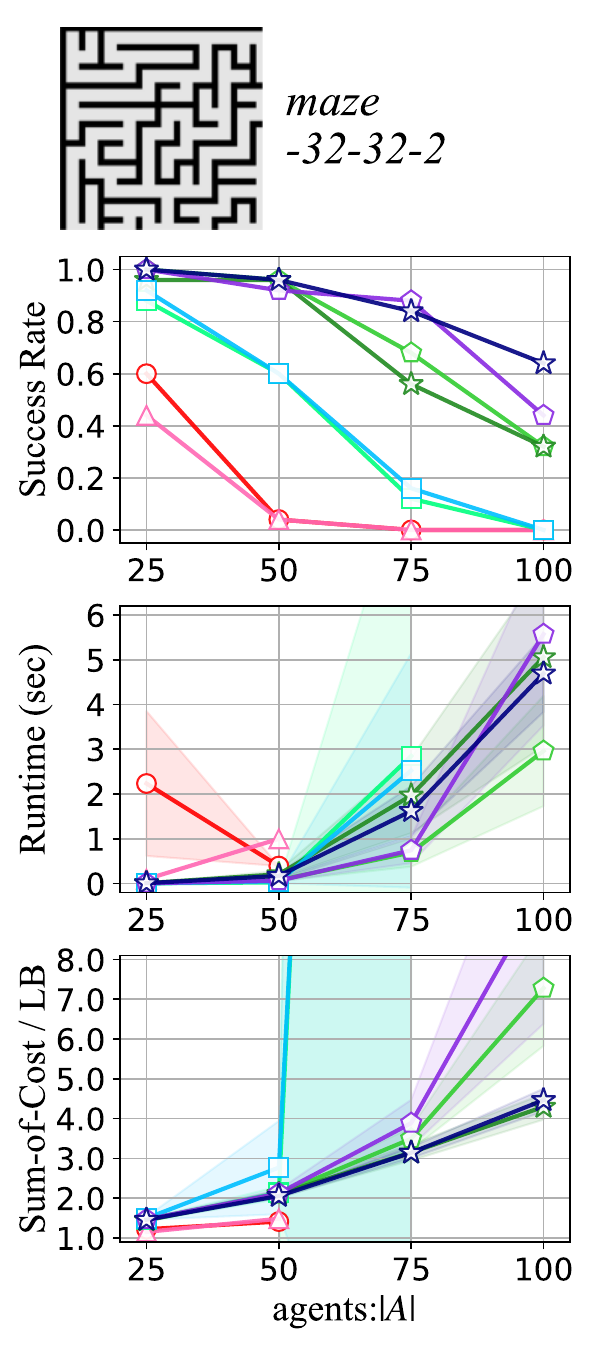}
  \entry{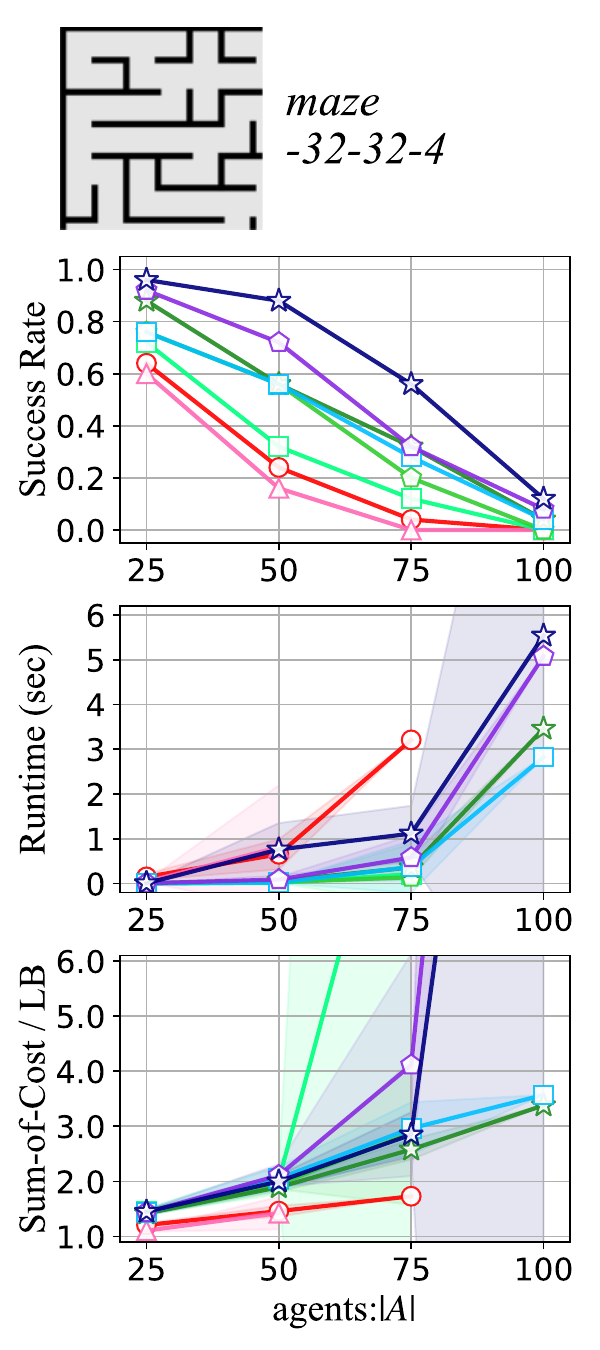}

  \entry{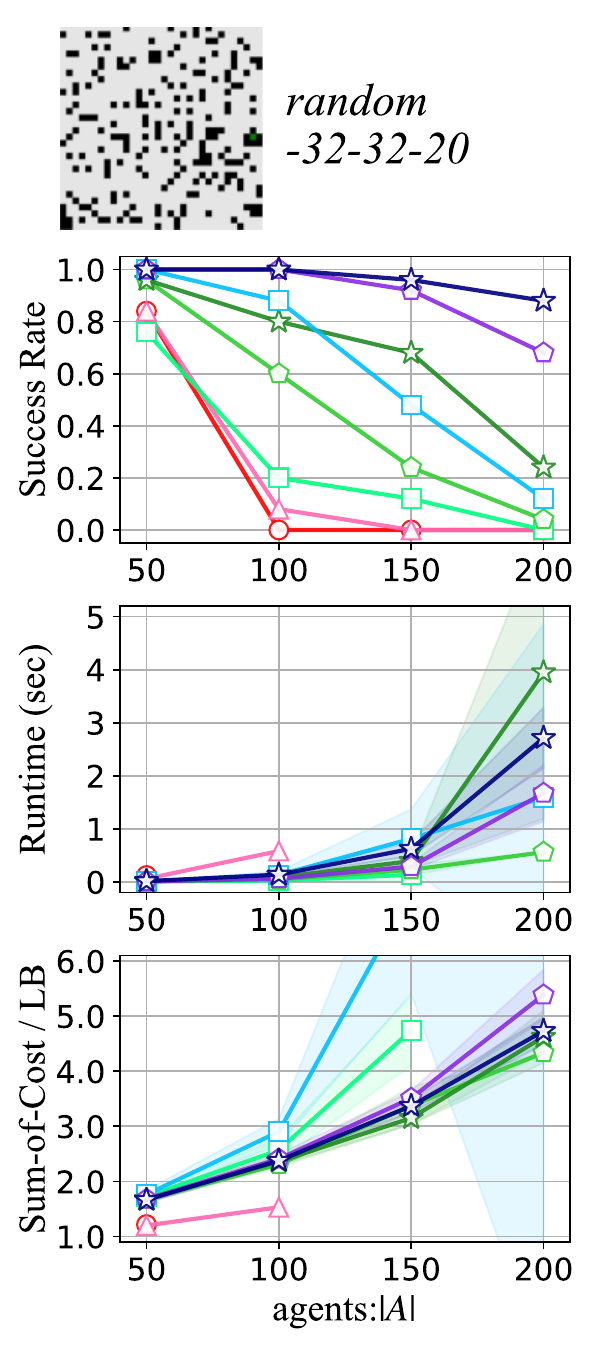}
  \entry{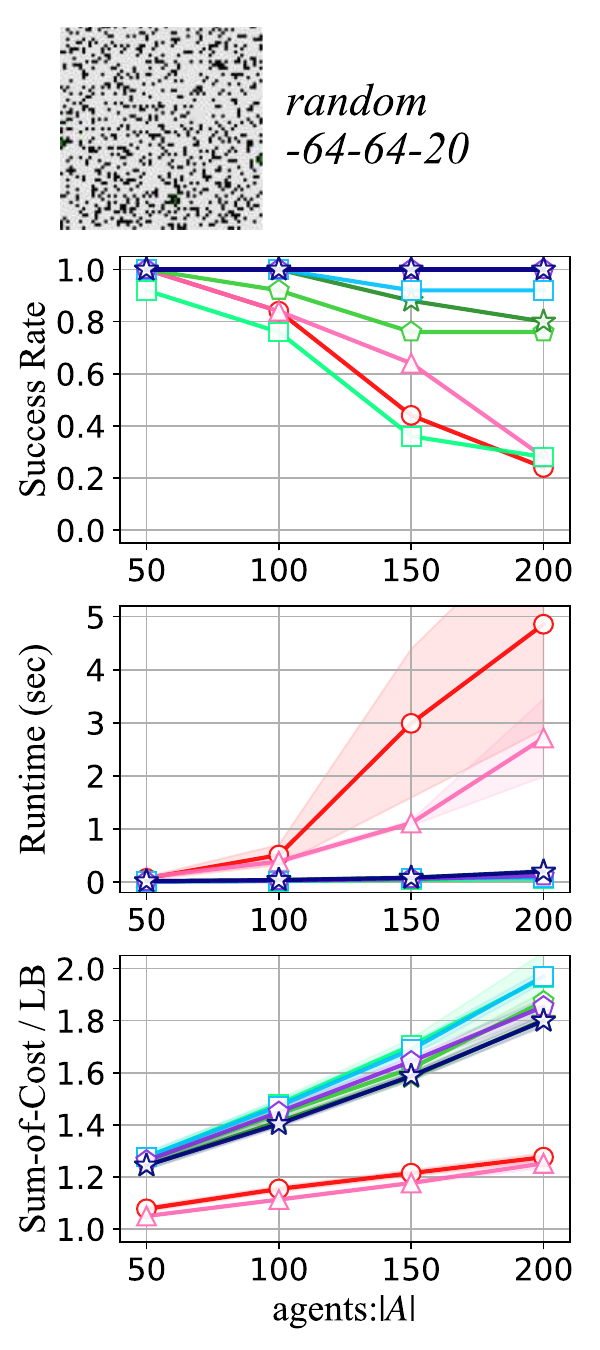}
  \entry{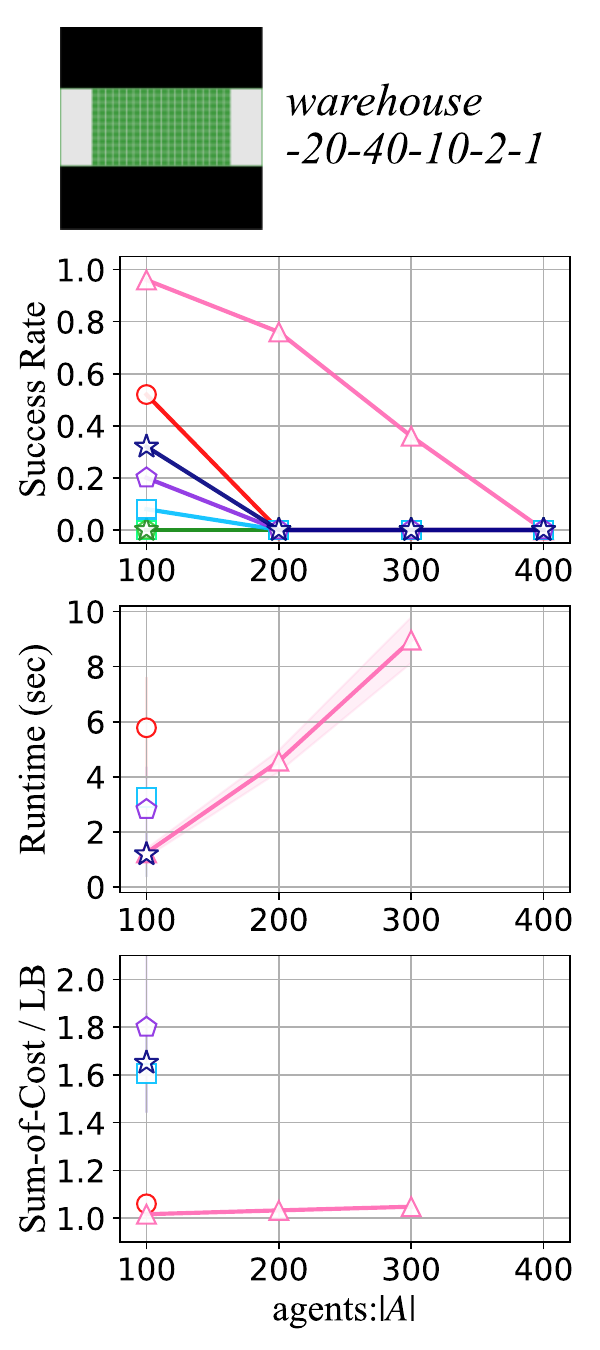}
  \entry{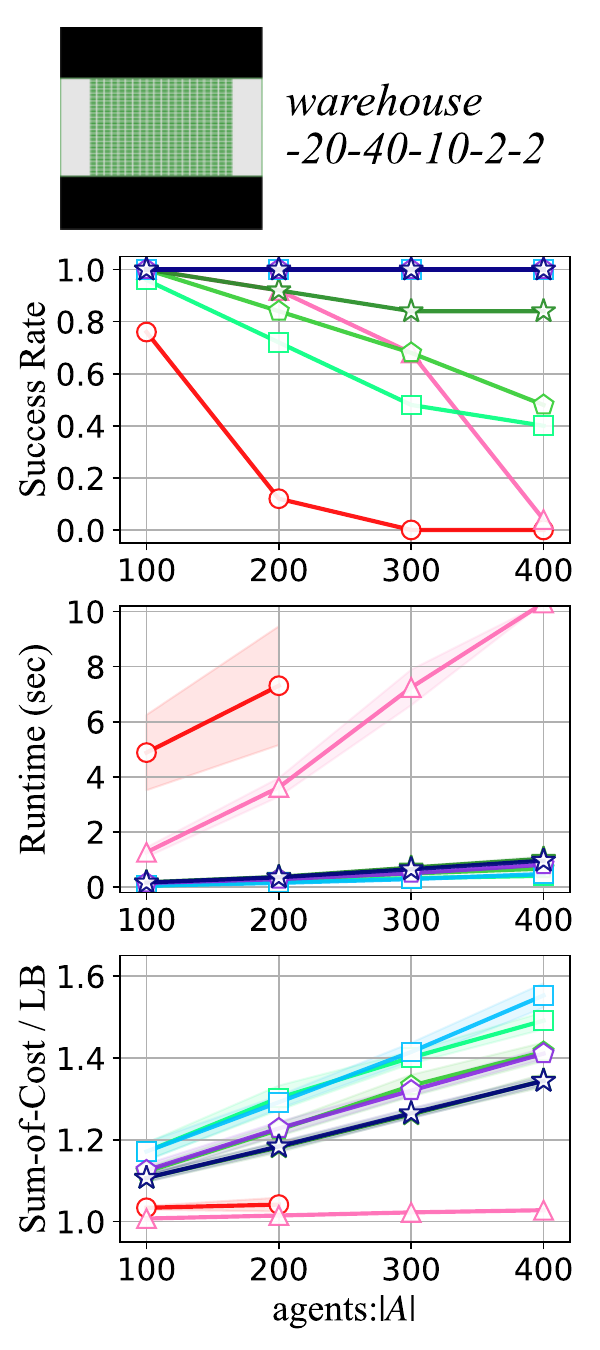}
  \entry{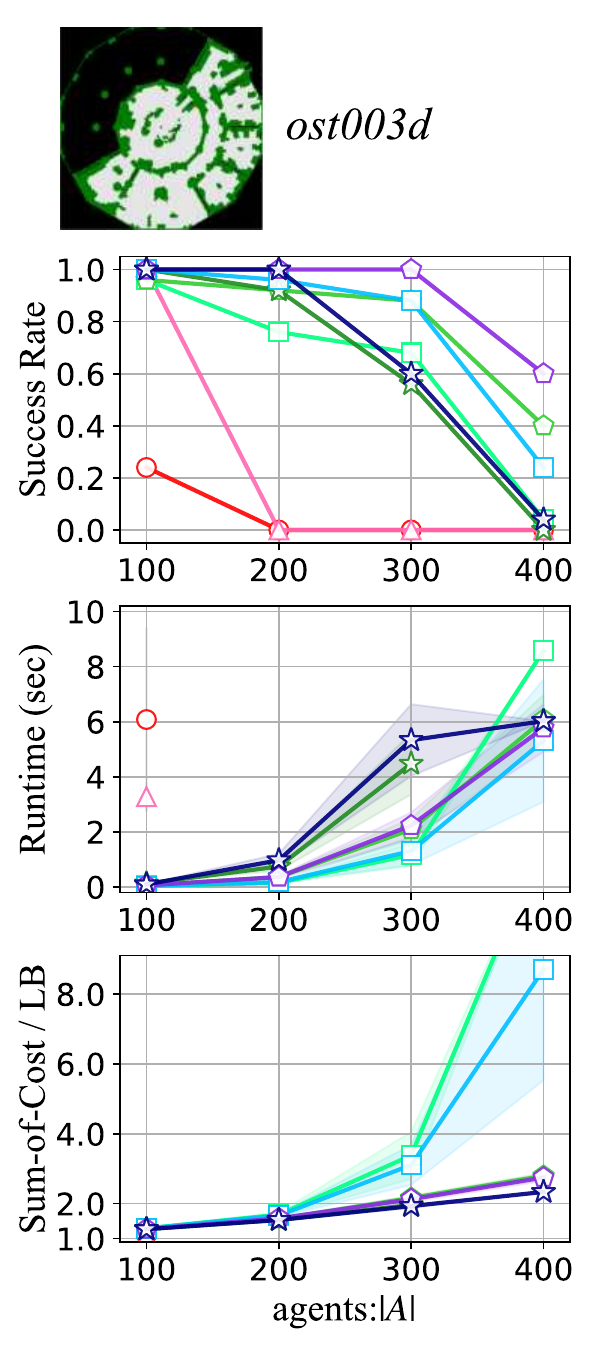}
  \entry{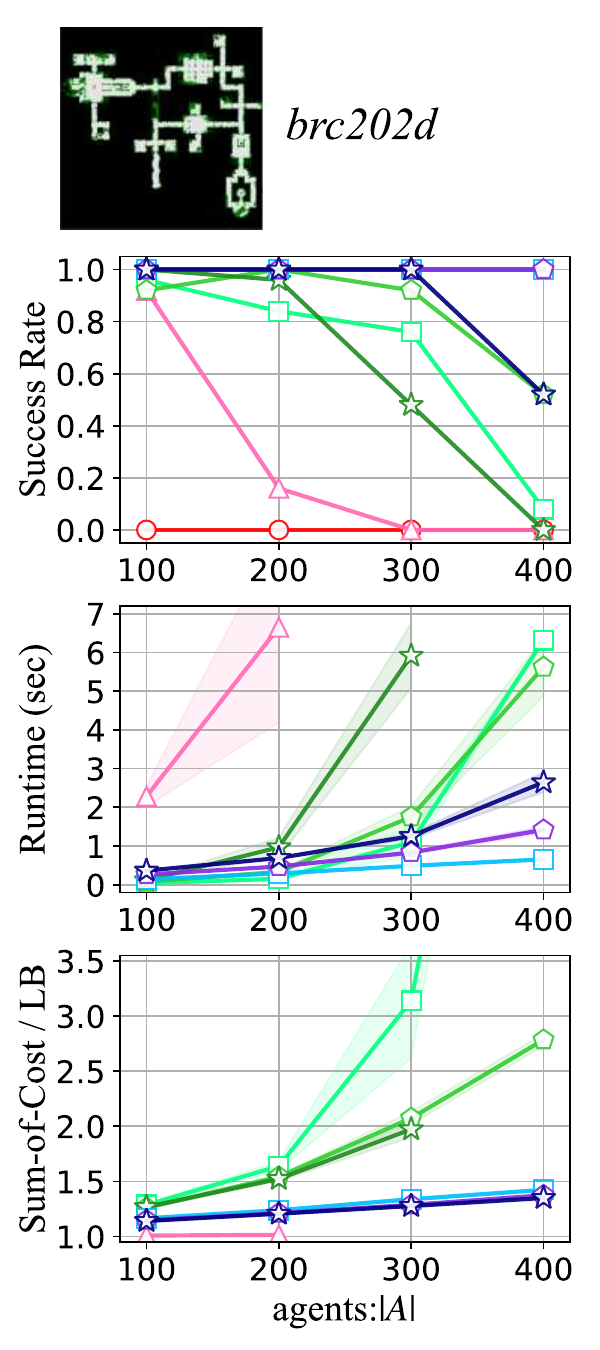}

  \includegraphics[width=1.0\linewidth]{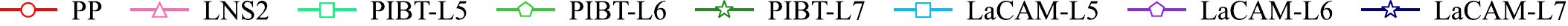}
  \caption{
    Results for one-shot MAPF.
    The success rate of planning within \SI{10}{\second} (top), average runtime (middle) and SoC normalized by lower bound ($LB =\sum_{i\in A} c_i^{\star}$, where $c_i^{\star}$ is the shortest-path cost of agent $i$ on the MAWPF configuration graph, from $(x^s_i, y^s_i, \theta_i^s, v{=}0)$ to $(x^g_i, y^g_i, \theta_i^g, v{=}0)$, ignoring all other agents; $1.0$ is minimum; bottom) for successful cases are shown. 
    $V_{\mathrm{max}}{=}2$ and $T_{\mathrm{rot}}{=}2$ were used.
  }
  \label{fig:oneshot-mapf}
\end{figure*}
}

{
  \begin{figure}[t!]
    \centering
    \includegraphics[width=1\linewidth]{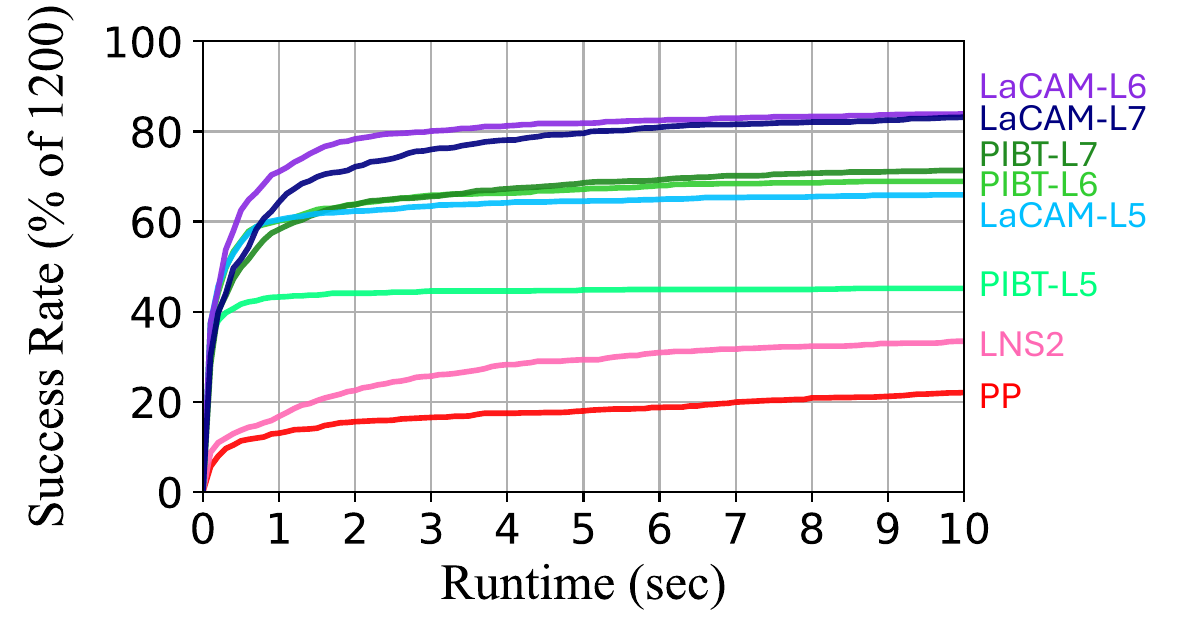}
    \caption{The
number of solved instances among 1,200 instances on 12 four-connected grid maps shown in \cref{fig:oneshot-mapf}.}
    \label{fig:all}
  \end{figure}
}

{
\newcommand{\entryon}[1]{
  \begin{minipage}{0.448\linewidth}
    \centering
    
    \includegraphics[width=1\linewidth,trim={0cm 0cm 0cm 0cm},clip]{fig/raw/#1_heatmap_on_large.pdf}
  \end{minipage}
}

{
\begin{figure*}[t!]
  \centering

  \entryon{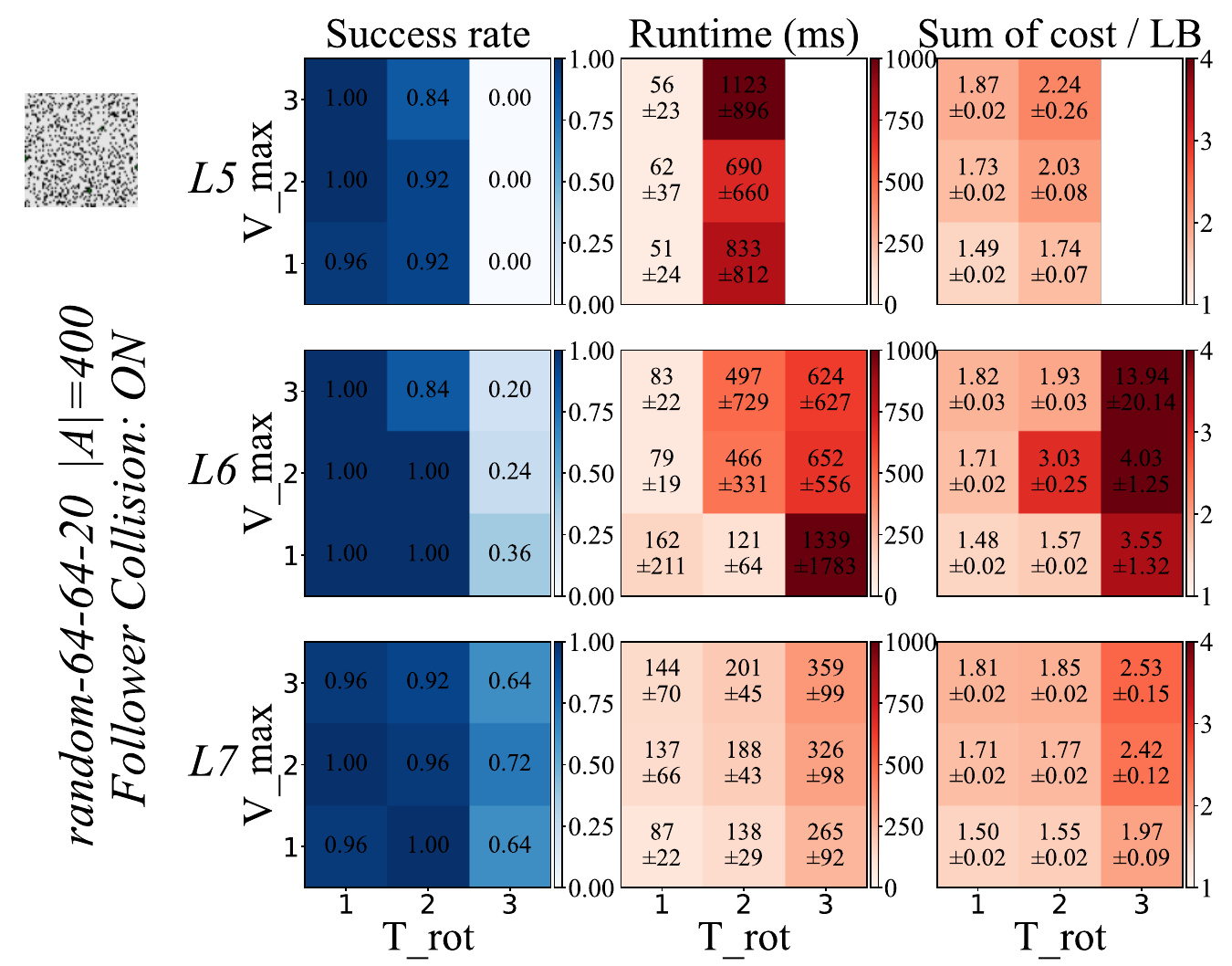}
  \entryon{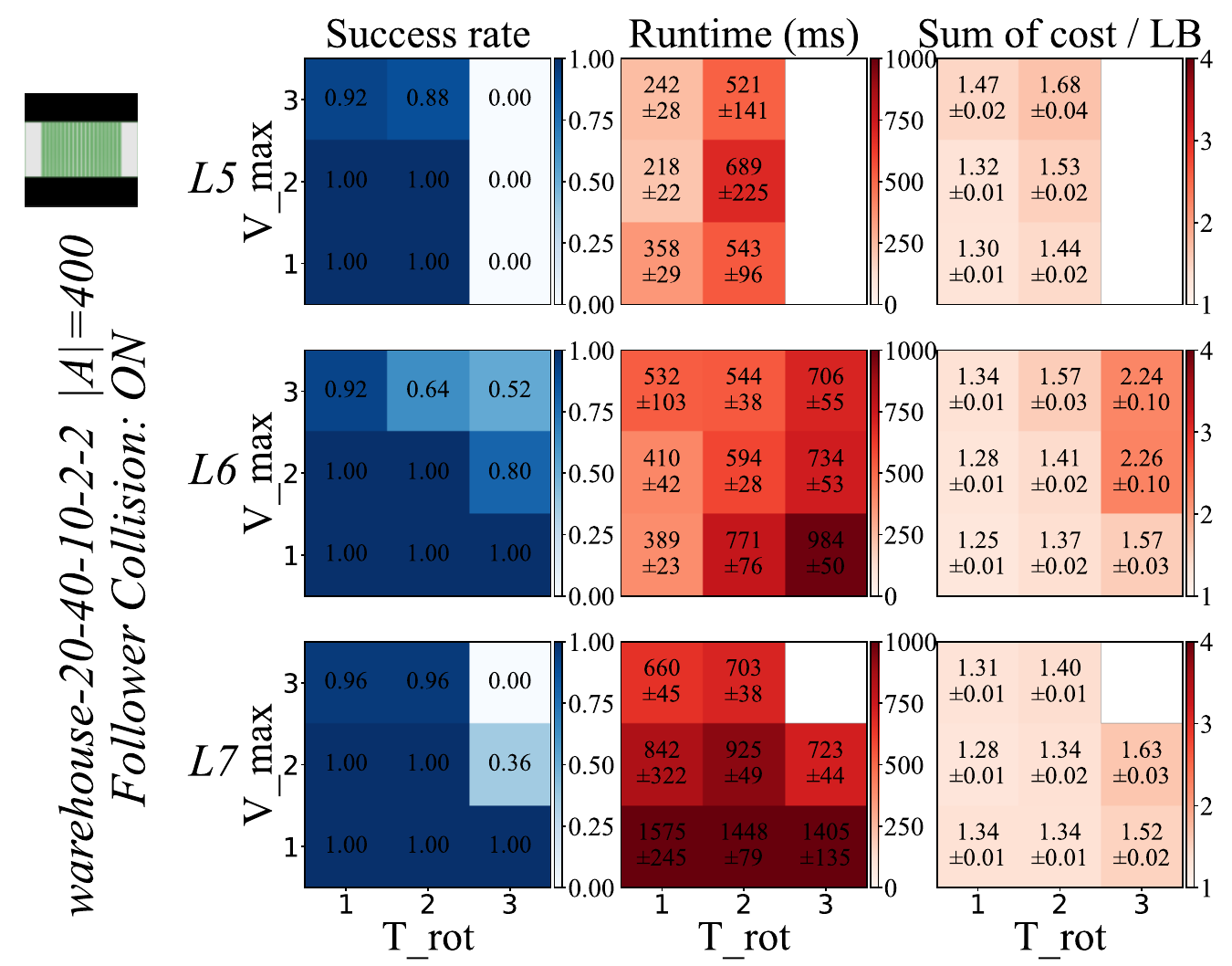}
  
  \caption{
    Results of experiments using LaCAM for MAWPF, varying the maximum speed and the number of steps required for a \SI{90}{\deg} turn.
    The success rate of planning within \SI{10}{\second} (left), average runtime (middle), and normalized SoC for successful cases are shown.
  }

  \label{fig:heatmaps}
\end{figure*}
}

\subsection{One-shot MAWPF Results}
\paragraph{Overview.}
The results, shown in \cref{fig:oneshot-mapf}, follow a trend similar to classical MAPF.
Across most maps, LaCAM achieved the highest success rate and consistently solved instances with several hundred agents within a few seconds.
This suggests that the LaCAM+PIBT pipeline preserves a certain degree of scalability and real-time performance in MAWPF despite richer kinodynamic constraints.
In contrast, PIBT alone yielded lower success rates, indicating that LaCAM is critical for handling difficult configurations.
PP and LNS2 showed lower success rates and generally longer runtimes than LaCAM.
Although performance depended on each map, PP and LNS2 typically succeeded on instances with about 50--100 agents, while LaCAM solved a substantially larger fraction within the same time budget.
\Cref{fig:all} summarizes the relationship between runtime and success rate across all scenarios and further highlights LaCAM's high success rate and responsiveness.

\paragraph{Anomaly on \emph{warehouse-20-40-10-2-1}.}
A notable exception is \emph{warehouse-20-40-10-2-1}, where LaCAM attains substantially lower success rates than PP and LNS2.
This is consistent with PIBT's sensitivity to narrow passages: width-one corridors often induce severe contention and blocking, making greedy, inheritance-based decisions prone to stalling~\cite{LaCAM2}.
This interpretation is supported by \emph{warehouse-20-40-10-2-2}: LaCAM solved all instances there but failed frequently on \emph{warehouse-20-40-10-2-1}, with corridor width (two vs.\ one) as the key difference.
Thus, the drop is a plausible consequence of applying PIBT to maps dominated by long, width-one corridors.

\paragraph{Solution Quality}
was broadly consistent with classical MAPF~\cite{LaCAM}. PP and LNS typically produced smaller sum-of-cost than LaCAM+PIBT.
This is expected because PP/LNS repeatedly compute (near-)shortest single-agent paths via A*-like searches, whereas PIBT is a feasibility-driven heuristic based on greedy ordering, and does not explicitly optimize objectives such as sum-of-cost.

\paragraph{Effects of Path Length in PIBT/LaCAM.}
Larger $L$ generally increased runtime because our PIBT adaptation employs a rolling-horizon interface: PIBT computes an $L$-step configuration array at each expansion, while LaCAM commits only the first configuration.
On many maps, $L{=}\{6, 7\}$ achieved higher success rates than $L=5$, suggesting that longer lookahead reduces dead-ends and failed successor generation.
It can also improve solution quality by mitigating locally feasible but globally inefficient choices that later require detours or waiting, thereby reducing sum-of-cost.

{
\newcommand{\Entry}[1]{
  \begin{minipage}{0.485\linewidth}
    \centering
    
    \includegraphics[width=1\linewidth,trim={0.3cm 0.4cm 0.5cm 0.5cm},clip]{fig/raw/result_fig3/ablation_#1.pdf}
  \end{minipage}
}

\begin{figure}[t!]
  \centering
  \Entry{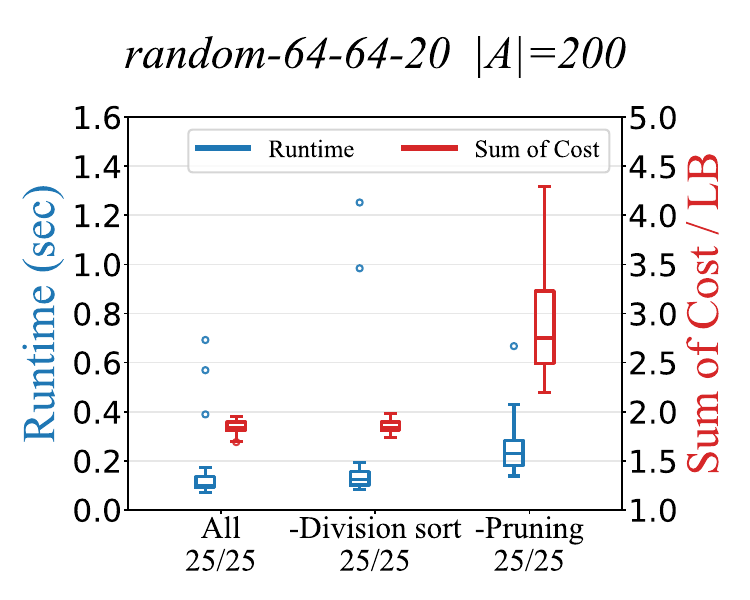}
  \Entry{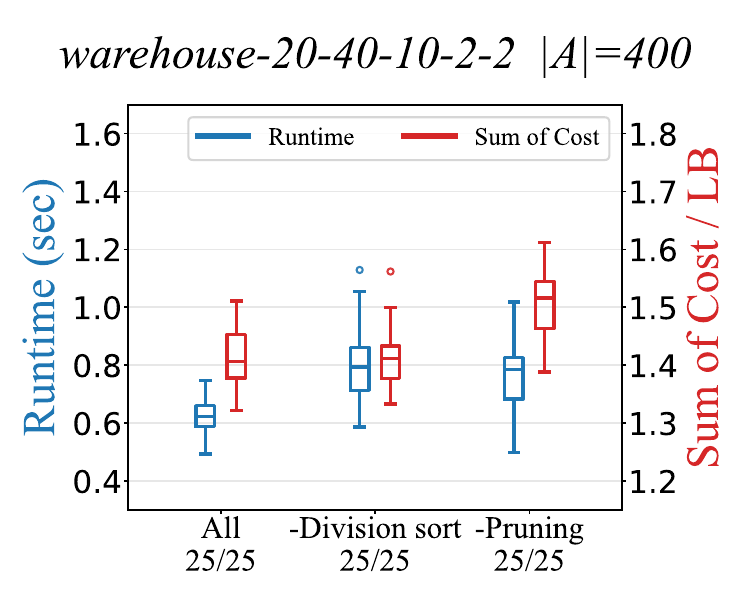}
  \caption{
    Results of the ablation study (LaCAM, $L=6$, $V_{\mathrm{max}}=T_{\mathrm{rot}}=2$).
    “All” denotes the full method, while ``-Division sort'' and ``-Pruning'' indicate ablated variants where each respective component is excluded.
    Planning was successful in all scenarios.
  }
  \label{fig:ablation}
\end{figure}
}

\subsection{Ablation Study}
To assess \emph{division sort} and \emph{pruning} for improving PIBT, we conducted an ablation study (\cref{fig:ablation}).
Division sort reduced runtime, indicating that ranking candidate paths is a non-trivial bottleneck, since PIBT for MAWPF generates far more candidates than the original PIBT under the enriched transition model.
Pruning further decreased runtime by removing candidates with identical endpoint configurations, and it improved sum-of-cost by filtering redundant detours such as oscillatory motions.

\subsection{Experiments with Other Problem Settings}
We further examined how kinodynamic parameters influence performance, focusing on LaCAM, which achieved the highest success rate among the compared methods. We fixed $L=6$ and varied $V_{\max}$ and $T_{\mathrm{rot}}$, summarizing success rate, runtime, and sum-of-cost in \cref{fig:heatmaps}.

Overall, larger $T_{\mathrm{rot}}$ degrades success rate and solution quality. Longer rotations consume more timesteps within the short horizon, leaving fewer steps for translation; thus, lookahead becomes less effective, and the planner behaves more like a smaller-$L$ setting, making feasibility harder to sustain and typically increasing sum-of-cost. 

The effect of $V_{\max}$ on runtime depends on map structure. On \emph{random-64-64-20}, runtime increases with $V_{\max}$, consistent with a larger search space. On \emph{warehouse-20-40-10-2-2}, higher $V_{\max}$ can reduce runtime, likely because long straight corridors better exploit acceleration and yield simpler, less congested progress, thereby reducing planning effort.
We also conducted experiments allowing follower collisions; details are provided in the appendix.

\section{Conclusion and Discussion}
We define multi-agent warehouse pathfinding (MAWPF), extending classical MAPF with differential-drive AGV constraints---multi-step rotations, acceleration/deceleration, and conservative collision definition to enhance safety.
Our empirical observation is that, among popular lightweight MAPF solvers, a rolling-horizon LaCAM+PIBT adaptation scales to hundreds of agents on benchmarks, achieving high success rates within a few seconds.

\paragraph{Why can't we scale to thousands of agents?} In classical MAPF, LaCAM can solve benchmark instances with on the order of thousands of agents. In MAWPF, while LaCAM was the most scalable among the compared methods, its practical limit in our experiments was only a few hundred agents. We attribute this gap to two factors. 

First, configuration generation is expensive. Each LaCAM expansion must produce a feasible next configuration, implemented in our MAWPF adaptation via a PIBT-style short-horizon procedure. Because MAWPF imposes richer kinodynamic constraints, even after pruning each agent can still have dozens to $\sim$100 candidate short paths, increasing the cost per configuration. 

Second, as discussed in Sec.~3, the MAWPF configuration graph is directed since heading/speed constraints make transitions hard to reverse. This encourages dead-end states and can slow LaCAM's high-level search. These observations are for one-shot MAWPF; lifelong MAWPF may handle more agents via frequent replanning and incremental progress.

\paragraph{Outlook.}
Future work includes improving solution quality and robustness on challenging layouts (e.g., narrow corridors) via more cost-aware successor generation and additional refinement mechanisms.

\section*{Acknowledgments}
This research was supported by a gift from Murata Machinery, Ltd.
\bibliographystyle{sty/named}
\bibliography{sty/ref-macro,ref}

\newpage

\clearpage
\onecolumn
\appendix

\section{Algorithm Details}
\label{app:algo_details}
The original PIBT algorithm and our MAWPF adaptation of LaCAM are summarized in \Cref{alg:pibt,alg:lacamMAWPF}, respectively.

%--------------------------------------
% Algorithm 2: PIBT
%--------------------------------------
\begin{algorithm}[]
  \caption{PIBT}
  \label{alg:pibt}
  \begin{algorithmic}[1]
  \small
    \Input{configuration $\Q\from$, agents $A$, goals $(g_i)_{i \in A}$}
    \Output{configuration $\Q\to$ (initially $\Q\to{[i]} = \bot$ for all $i \in A$)}

    \For{$i \in A$}
      \If{$\Q\to[i] = \bot$}
        \State $\PIBT(i)$
      \EndIf
    \EndFor
    \State \Return $\Q\to$

    \Procedure{\PIBT}{$i$} 
      \State $C \gets \neigh(\Q\from[i]) \cup \{\,\Q\from[i]\,\}$ 
      \State sort $C$ in ascending order of $\dist(u, g_i)$ for $u \in C$

      \For{$v \in C$}
        \If{assigning $v$ to agent $i$ in $\Q\to$ would cause a collision}
          \State \Continue
        \EndIf

        \State $\Q\to[i] \gets v$

        \If{$\exists j \in A$ such that $j \neq i \land \Q\from[j] = v \land \Q\to[j] = \bot$}
          \If{$\PIBT(j) = \invalid$}
            \State \Continue
          \EndIf
        \EndIf

        \State \Return \valid
      \EndFor

      \State $\Q\to[i] \gets \Q\from[i]$
      \State \Return \invalid
    \EndProcedure
  \end{algorithmic}
\end{algorithm}

%--------------------------------------
% Algorithm 1: LaCAM
%--------------------------------------
\begin{algorithm}[H]
  \caption{LaCAM for MAWPF}
  \label{alg:lacamMAWPF}
  \begin{algorithmic}[1]
  \small
    \Input{MAPF instance}
    \Output{a solution or \failure}

    \State initialize \mapname{Open} as a stack
    \State initialize \mapname{Explored} as a hash table mapping configurations to nodes
    \State $\N\init \gets \{\,\mapname{config}: S,\; \mapname{tree}: \langle \C\init \rangle,\; \mapname{parent}: \bot\,\}$
    \State $\push(\mapname{Open}, \N\init)$; \quad $\mapname{Explored}[S] \gets \N\init$

    \While{$\mapname{Open} \neq \emptyset$}
      \State $\N \gets \funcname{top}(\mapname{Open})$
      \If{$\N.\mapname{config} = \m{\mathcal{G}}$}\Comment{goal}
        \State \Return \funcname{backtrack}$(\N)$
      \EndIf

      \If{$\N.\mapname{tree} = \emptyset$}
        \State $\pop(\mapname{Open})$ \Comment{discard this search node}
        \State \Continue
      \EndIf

      \State $\C \gets$ next element removed from $\N.\mapname{tree}$ \Comment{take one constraint}
      \State \funcname{low\_level\_expansion}$(\N, \C)$
      \State $\diff{\{\,\Q\to_0,...,\Q\to_{L-1}\,\}} \gets \funcname{configuration\_generator}(\N, \C)$
      \State $\diff{\Q\new \gets \Q\to_0}$\Comment{Rolling horizon method}

      \If{$\Q\new = \bot$}; \Continue \Comment{generator may fail}
      \EndIf
      \If{$\mapname{Explored}[\Q\new] \neq \bot$};\Continue
      \EndIf

      \State $\N\new \gets \{\,\mapname{config}: \Q\new,\; \mapname{tree}: \langle \C\init \rangle,\; \mapname{parent}: \N\,\}$
      \State $\push(\mapname{Open}, \N\new)$
      \State $\mapname{Explored}[\Q\new] \gets \N\new$
    \EndWhile

    \State \Return \failure  \Comment{no solution exists}
  \end{algorithmic}
\end{algorithm}

\clearpage
\section{Influence of Follower Collision}
\label{app:follower_collision}

To examine the impact of the follower-collision constraint, we repeated the same experimental protocol as in Sec.~6.4 while allowing follower collisions.
The results are summarized in \Cref{fig:heatmaps2}.
Contrary to our initial expectation, enforcing the follower-collision constraint yields a higher success rate.
We conjecture that prohibiting follower collisions acts as an implicit congestion-avoidance mechanism: it discourages tailgating behaviors that create tightly packed queues, thereby reducing the likelihood of entering irreversible ``deadlock'' states.

{
\newcommand{\entryoff}[1]{
  \begin{minipage}{0.44\linewidth}
    \centering
    
    \includegraphics[width=1\linewidth,trim={0cm 0cm 0cm 0cm},clip]{fig/raw/#1_heatmap_off_large.pdf}
  \end{minipage}
}

{
\begin{figure*}[t]
  \centering

  \entryoff{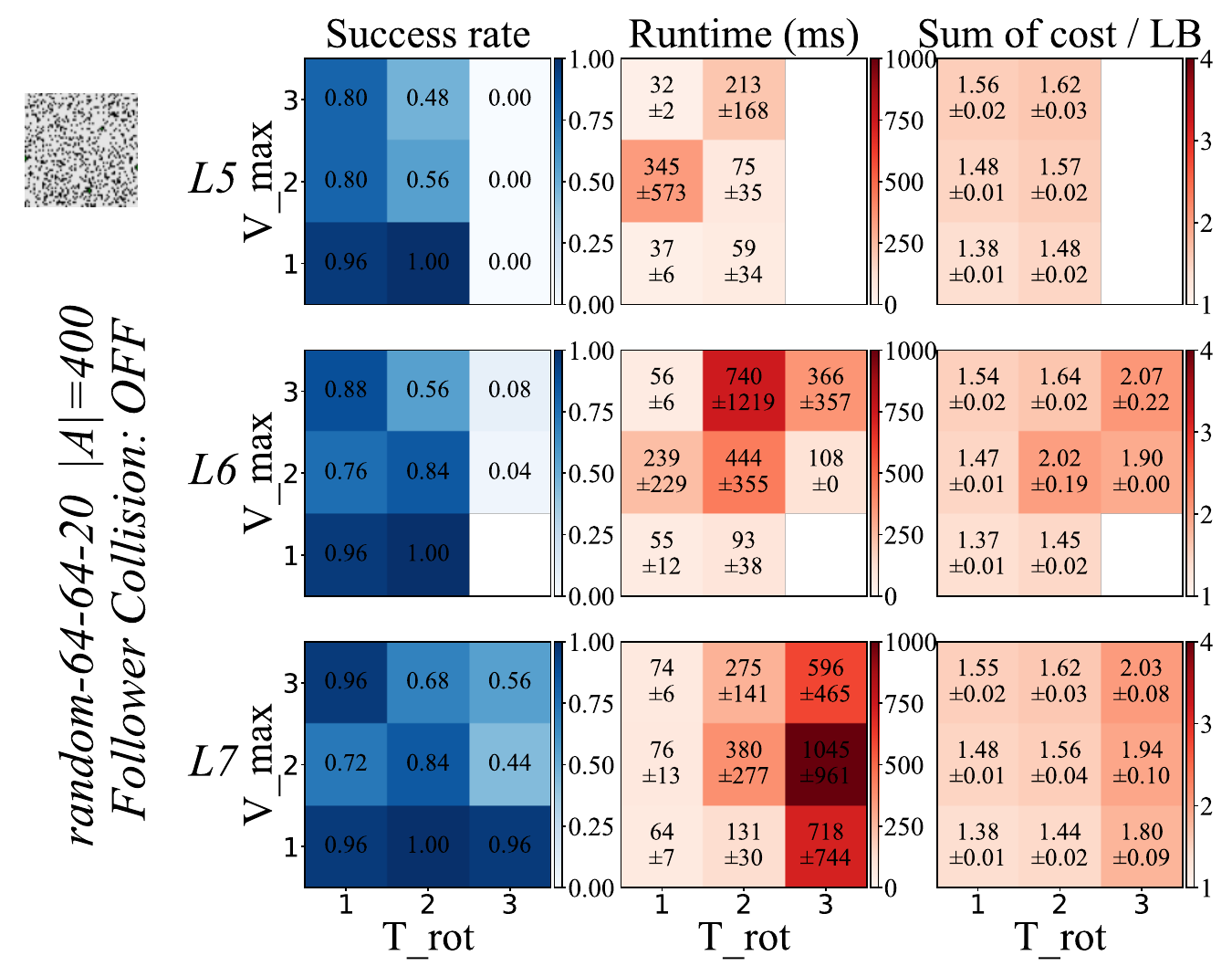}
  \entryoff{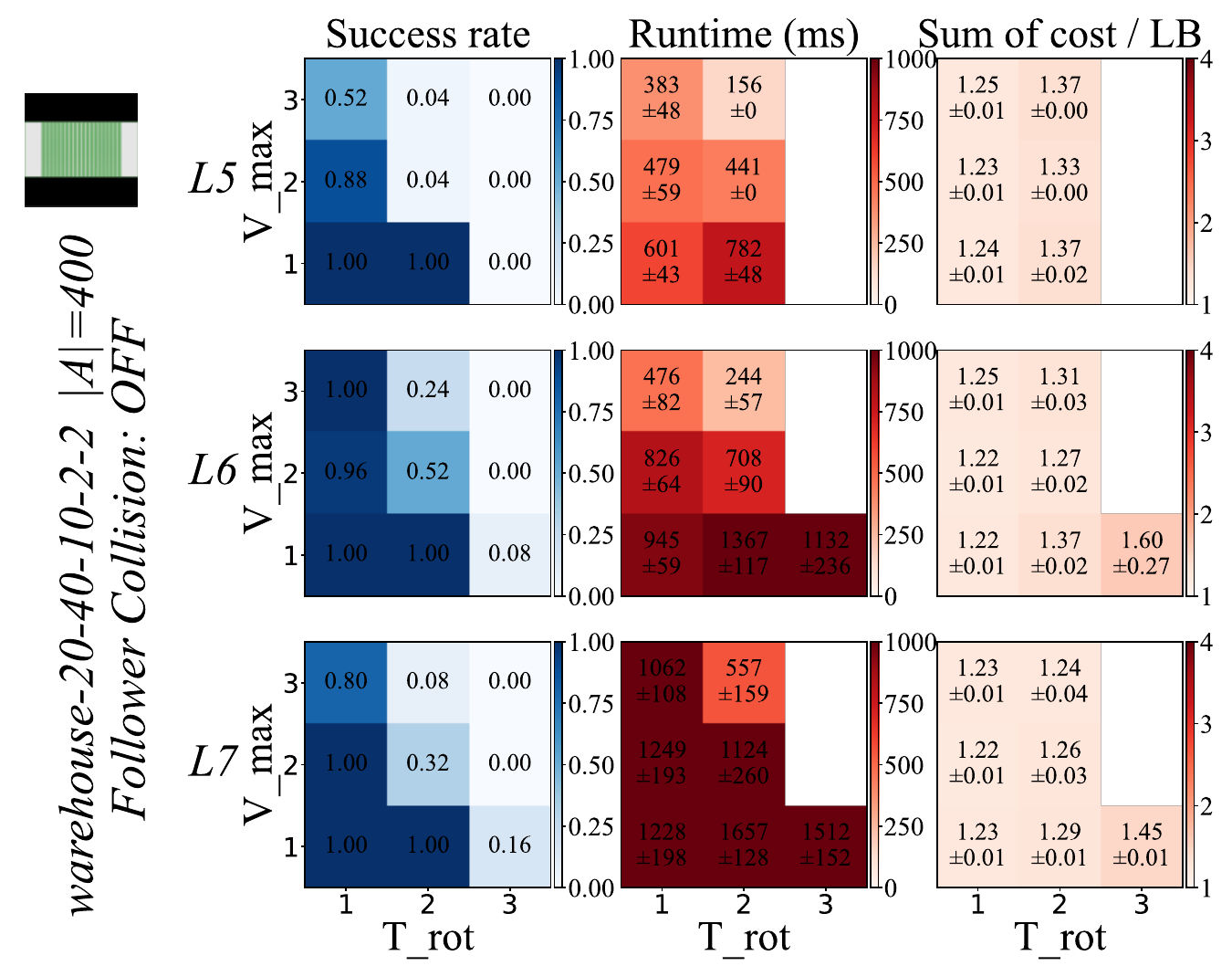}

  \caption{
  Results of experiments using LaCAM for MAWPF with follower collisions allowed, varying the maximum speed and the number of steps required for a \SI{90}{\deg} turn.
  The success rate of planning within \SI{10}{\second} (left), average runtime (middle) and SoC normalised by lower bound ($1.0$ is minimum; right) for successful cases are shown.
}

  \label{fig:heatmaps2}
\end{figure*}
}

\end{document}